# Is it possible to predict long-term success with k-NN? Case Study of four market indices (FTSE100, DAX, HANGSENG, NASDAQ)

**Y Shi, A N Gorban, T Y Yang**

Department of Mathematics, University of Leicester, LE1 7RH, UK

ys98@leicester.ac.uk , ag153@leicester.ac.uk, yty725@sina.com

**Abstract**. This case study tests the possibility of prediction for 'success' (or 'winner') components of four stock & shares market indices in a time period of three years from 02-Jul-2009 to 29-Jun-2012.We compare their performance ain two time frames: initial frame three months at the beginning (02/06/2009-30/09/2009) and the final three month frame (02/04/2012-29/06/2012).To label the components, average price ratio between two time frames in descending order is computed. The average price ratio is defined as the ratio between the mean prices of the beginning and final time period. The 'winner' components are referred to the top one third of total components in the same order as average price ratio it means the mean price of final time period is relatively higher than the beginning time period. The 'loser' components are referred to the last one third of total components in the same order as they have higher mean prices of beginning time period. We analyse, is there any information about the winner-looser separation in the initial fragments of the daily closing prices log-returns time series. The Leave-One-Out Cross-Validation with k-NN algorithm is applied on the daily log-return of components using a distance and proximity in the experiment. By looking at the error analysis, it shows that for HANGSENG and DAX index, there are clear signs of possibility to evaluate the probability of long-term success. The correlation distance matrix histograms and 2-D/3-D elastic maps generated from ViDaExpert show that the 'winner' components are closer to each other and 'winner'/'loser' components are separable on elastic maps for HANGSENG and DAX index while for the negative possibility indices, there is no sign of separation.

## 1. Introduction
Prediction of time series is essential and difficult task in the real world. Many methods had been studied these days such as constructing complex models for simulating the prices, different types of regression models and so on. It is important to study the predictability of the time series separately from constructing the models because in creation of each model we assume some additional hypotheses about the model structure. It is also interesting to study this problem as when there is a positive predictability or sign of changes in the time series, traders may use this as a sign to discover crisis to prepare a response for some critical situations. For example, during crisis, the correlation and variance is higher [1], [2].

In this work, we test the possibility of prediction of long-term success on the financial market. The time interval is there years. We evaluate the success of the companies during these three years. The main question is: was there a similarity between the most successful companies and dissimilarity between them and the least successful ones at the beginning of this time period? In other words, is there anything in the previous history that may give us information about the success in the following three years? We use movement of prices only and our study should demonstrate the possibility to use historical data for the long term success prediction.

This question is in the focus of the discussion between technical analysis and efficient market hypothesis. Technical analysis is applied by many traders and investors to discover trends and patterns and sometimes to assist them to construct efficient trading strategy. The Efficient Market Hypothesis (EMH) is the important assumption in the research study for technical analysis [3]. This hypothesis assumes there are no systematic biases in prices and therefore technical analysis could not generate consistent profit. The study on the results of simulating four technical trading rules over the 1960 to 1983 period indicate that it is too difficult for technical analysis to forecast subsequent prices [4].Filter rules constructed from the technical analysis are used on intra-daily foreign exchange market. These rules can generate some profits but not for the general case [5]. Many survey studies about profitability, theories and empirical work regarding technical trading strategies are reviewed but they cannot agree a consistent profitability of technical analysis [6]. The positive profitability of technical analysis could imply that historical prices contain information about future prices. Another interesting analysis of 20 new equity markets in emerging economies shows that the correlation is lower for the developed country returns [7]. This could link to the result of HANGSENG index which is the only developing country index.

Our case study is aimed to find possibility of the prediction of 'success' within three years' time interval from 02-JUL-2009 to 29-JUN-2012 for four selected stock and share market indices. We compare their performance in two time frames: initial frame three months at the beginning (02/07/2009-30/09/2009) and the final three month frame (02/04/2012-29/06/2012).The idea of the main experiment is based on backward analysis. The backward analysis can be defined as an analysis to determine properties of the inputs of a program from properties or contexts of outputs. This case study is aimed to construct experiments on the data to test if it is possible to predict the long-term success. The possibility of long-term 'success' of the selected indices is tested from the results of the experiment in the three years' time interval .For each stock market index, the closing prices of all components are collected from Yahoo! Finance website. After data pre-processing step, the remaining components are labeled with 'winner' companies or 'loser' companies (or simply just 'winner', 'loser') by using the '1/3 average price' approach. For this approach, we compute the average price ratio which is defined as the mean price of the end period divided by the mean price of beginning period. The companies are then sorted by the descending order of this computed ratio. We label the first 33.3% of companies as 'winner' and label the last 33.3% of the companies as 'loser'. Then the log-return prices are computed on this data. The k-NN algorithm with Leave-One-Out Cross-Validation with two distance measurements is used as the indicator to test the possibility of prediction.

We use Leave-One-Out Cross-Validation for k-NN classifier to test the possibility of prediction of 'success' components for each market index. The data is collected from data and cleaned. Then we did experiment of Cross-Validation for k-NN using two different forms of distance measurements. Then we analyze the total error and separate error and use two methods to visualize our result. We investigate that there is a possibility of predictions for long-term success for HANGSENG and DAX indices in the result section. We summarize our result and conclude in the last part.

## 2. Methods and backgrounds
### 2.1. Data Pre-processing
The closing price is the final price at which a security (in this case study, the stock exchange) is traded on a given trading day. It represents the value of this security on a trading day until it changes again on the next trading day. The raw data used for the experiments of this paper are the closing prices of all components for each index for different time frames. (3 months, 6 months, 9 months, 12 months, 15 months and 18 months) These closing prices are collected from Yahoo! Finance. The first cleaning step is to compare the dates of each company with the index's date. The prices on the dates that are not in the trading date of the index are deleted. The second step is dealing with the missing values. For companies have more than 20% of missing values are deleted from list of companies for further experiment. For the rest of companies with missing values, the missing values are filled with the attribute mean (mean of the closing price on specific date). The closing prices are sorted from the oldest to nearest and they are saved in a matrix with each column represents each company and each row represents each date.

### 2.2. Log Returns
The k-NN algorithm is applied using the next day's daily log-return of closing prices. There are two main reasons for using returns. First, for average investors, return of an asset is a complete and scale-free summary of the investment opportunity. Second, return series are easier to handle than prices series as they have more attractive statistical properties [8]. Let $P_t$ be the closing price of each company at time t. The log-return at time t is defined as:

$$r_t = ln\frac{P_t}{P_{t-1}}.$$

Hence if consider the log-returns as a matrix defined as the form of closing prices, the matrix of log returns will have one less row than the matrix of closing prices. The advantages of log returns over the simple net returns are obvious. First, the log return is the sum of continuously compounded one-period returns involved. Second, statistical properties of log returns are more tractable.

### 2.3. Pearson's Product-moment Coefficient
Two functions of correlation coefficients are the measurements of distance when applying k-NN algorithm in the case study. The correlation coefficient is an indicator of measuring the dependence of two random variables. The Pearson's Correlation or Pearson's Product-moment coefficient is defined as the covariance of two random variables divided by the product of the individual standard deviations. For a series of n measurements of $X$ and $Y$ written as $x_i$ and $y_i$, where $i = 1,2,...,n$. The correlation coefficient is defined as:

$$r_{xy} = \frac{\sum_{i=1}^{n}(x_i-\bar{x})(y_i-\bar{y})}{\sqrt{\sum_{i=1}^{n}(x_i-\bar{x})^2 \sum_{i=1}^{n}(y_i-\bar{y})^2}}.$$

The coefficient is bounded between +1 and -1. When coefficient equals to 0, it means no linear relationship is found between $X$ and $Y$. When coefficient equals to +1, it means there is a positive linear relationship between $X$ and $Y$. When coefficient equals to -1, it means there is a negative linear relationship between $X$ and $Y$.

### 2.4. Correlation Distance
In cluster analysis, the correlation distance is used in a specific metric. The correlation distance on returns of time series $(X,Y)$ is defined as:

$$d(X,Y) = \sqrt{2(1-c_{XY})}, \qquad \text{(Distance)}$$

where $X = (x_1, x_2, ..., x_T)$, $Y = (y_1, y_2, ..., y_T)$ and $c_{XY}$ is the correlation coefficient between $X$ and $Y$[9]. It is used as in the analysis of a case study for the Italian Hospitality Sector [10]. Proximity of this distance is computed below.

$$d(X,Y) = \sqrt{1 - c_{XY}^2}, \tag{Proximity}$$

where $X = (x_1, x_2, ..., x_T)$, $Y = (y_1, y_2, ..., y_T)$ and $c_{XY}$ is the correlation coefficient between $X$ and $Y$. Consider the geometric interpretation of the correlation coefficient, it can be thought as the cosine with an angel between $x_i - \bar{x}$ and $y_i - \bar{y}$ where $x_i \in X$ and $\bar{x}$ is the mean of all $x_i$, $y_i \in Y$ and $\bar{y}$ is the mean of all $y_i$. Using double formulas of cosine, the distance function can take another expression $2\sin\alpha$ where $\alpha \in [0, \frac{\pi}{2}]$.

$$2\sin\alpha = \sqrt{2(1 - \cos 2\alpha)} \tag{2.1}$$

The distance $d(X,Y)$ can have another expression of $2\sin\alpha$. The proximity is defined in the similarly way, with an expression of $\sin(2\alpha)$.

$$\sin(2\alpha) = \sqrt{1 - \cos^2(2\alpha)} \tag{2.2}$$

This expression is 0 when $\alpha = 0$ and $\alpha = \frac{\pi}{2}$.

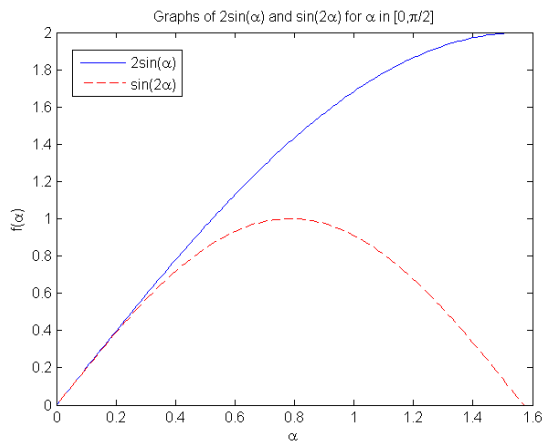

[Figure 2.1] Graph of $2\sin(\alpha)$ (Distance) and $\sin(2\alpha)$ (Proximity of Distance) for $\alpha \in [0, \frac{\pi}{2}]$.

The comparison of these two expressions is shown in [Figure 2.1]. These two expressions are the same when $\alpha$ is small and grows linearly as $\alpha$ increases. But the plot of proximity starts to reach its maximum at 1 when $\alpha$ is approaching to 0.8 approximately and then it starts to decreases and reaches zero again at around $\alpha = 1.6$. This small difference has almost no effect on the later experiments.

*2.5. Average Price and Long term success*

Before applying the k-NN algorithm, the training data (data with labels) are required. The '1/3 average method' is used to label the components (i.e. label them with either 'winner' or 'loser'). The 'winner' companies are the companies has relatively higher average price in the last time frame while the 'loser' companies are the ones have relatively higher average price in the beginning time frame. The average price of the time series in the specific time period is defined as the sum of the total prices in this period divided by the number of prices. (i.e. if in a time period there are $n$ prices, each specific price in this period is denoted as $P_1, P_2, \ldots, P_n$, the average price $Avp = \frac{1}{n}\sum_{i=1}^{n} P_i$.)

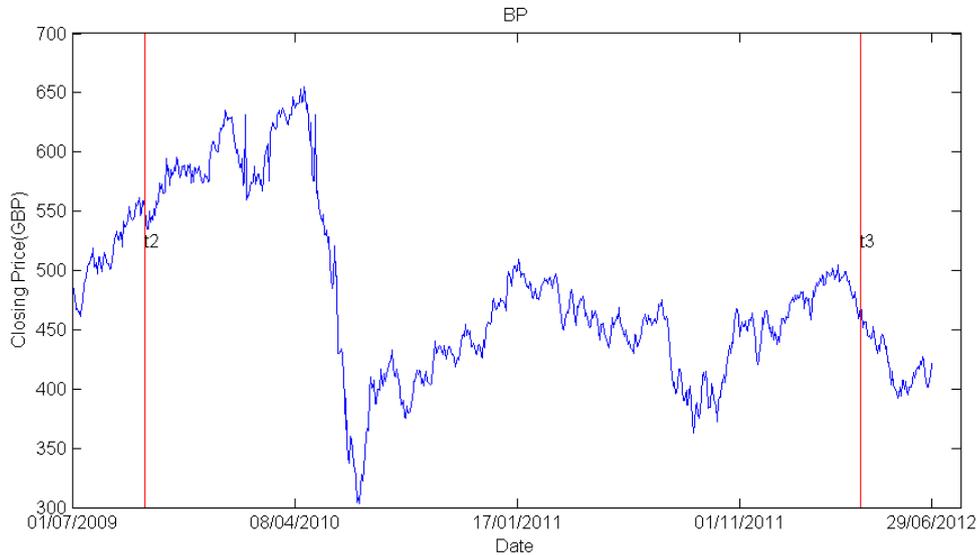

[Figure 2.2] Figure of company BP, component of FTSE100 index with date against its closing price in GBP between 01/07/2009 and 29/06/2012.

For example, [Figure 2.2] is a time series of a component of FTSE100 index. The red lines labeled with 't2' and 't3' represent the two border lines of the beginning and last 3 month time period. The average prices of the first and last time period is calculated using the definition introduced earlier and let $Avp_1$ be the average price for the first three months, and let $Avp_2$ be the average price for the last three months. The ratio of average price of this company is then defined as the ratio of $Avp_1$ and $Avp_2$. (i.e. $ratio = \frac{Avp_1}{Avp_2}$) The next step of '1/3 average price' method is to sort this ratio in descending order. The first 1/3 companies are labeled as 'winner' while the last 1/3 companies are labeled as 'loser'.

The Nearest-Neighbors (NN) predictors had been studied in several papers. The NN predictor is applied for the analysis on forecasting daily exchange data in foreign exchange markets [11]. This paper gives an interesting application of NN in the financial time series. An extension of similar study of using Simultaneous Nearest-Neighbor (SNN) predictors is applied to nine EMS currencies using daily data [12]. Hence it is interesting to use k-Nearest Neighbor as an indicator to test predictability.

*2.6. k-Nearest-Neighbor Algorithm and Leave-One-Out Cross-Validation*
The k-Nearest-Neighbor Classifier is one of the famous and simplest classification algorithms. It requires no models to fit [7]. The classification rule is for a specific test point with no label, it can be classified using majority vote among the k-Nearest Neighbors. The Nearest Neighbors of a test point are found by looking for the k smallest distances between the test point and the training points. The choice of distance functions has many forms. One of them is the Euclidean distance in the feature space:

$$d_{(i)} = \|x_{(i)} - x_0\|$$

where $x_0$ is a test point and $x_{(i)}$ are points in the training set. In this paper, the distances used will be two distances introduced earlier. To estimate the prediction error, cross-validation is used. This is probably the simplest and widely used method when there is no sufficient enough amounts of data [13].This predictor is also The K-fold cross-validation uses part of the data to fit the model, in my experiment, the k-NN classifier, and a different part to test it. In K-fold cross-validation, the data is split into K equal-sized parts. For the Kth part, when using model to predict the Kth part of data, the K-1 parts of data are used to fit the model. The Leave-One-Out Cross-Validation is a special case when K is the number of total number of data points. When applying this method to k-NN, it extracts one point from the original data and this point is considered as test point. Then the rest of them are used as the training points.

The time period of the experiments is chosen between 01/07/2009 and 29/06/2012. (For some stock markets, it may vary as it might not have prices at 01/07/2009 or 29/06/2012, and the next trading date will be counted as the boundary)Because of the financial crisis occurred in 2007/2008, the stock and exchange indices time series were affected.

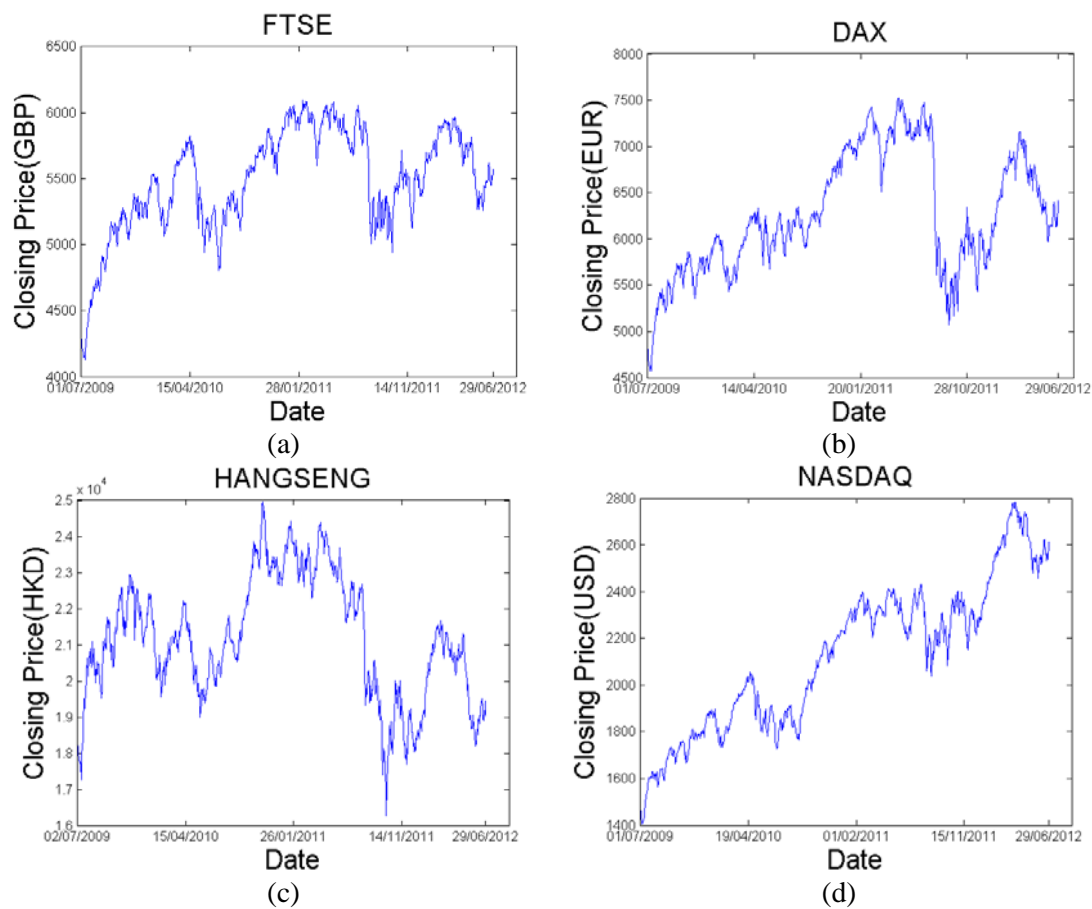

[Figure 2.3] Figures of market index in the three years period between 01/07/2009 and 29/06/2012 (For HANGSENG index, it does not have price on 01/07/2009 hence it starts from the next trading date) (a) Closing price of FTSE (b) Closing price of DAX (c) Closing price of HANGSENG (d) Closing price of NASDAQ

[Figure 2.3] shows the movement of the closing prices of four market indices between 01/07/2009 and 29/06/2012.The closing prices are relatively low in the beginning for all markets. Then they begin to increase and except for HANGSENG index, they have a generally up trend. This uptrend stands for the recovery of the market from crisis. For HANGSENG index, the closing price was increasing then at some date, it starts the downward trend and it was likely to be hit by the second wave of crisis.

*2.7. Estimate of Proportion*

In statistics, sampling theory studies the relationship between the population and sample from this population. Estimate of proportion is considered as an interval estimate of the population proportion [14]. Suppose that a population is infinite and the probability of occurrence of an event is $p$, and consider all possible samples of size N drawn from this population, a sampling distribution of proportions with mean µ and standard deviation σ is given by

$$\mu = p \text{ and } \sigma = \sqrt{\frac{p(1-p)}{N}}$$

In the error analysis part, the sampling distributions of 'winner' and 'loser' companies are computed. It gives more information in the case if the testing number of companies is small.

## 3. Results and analysis

The experiment is computed using MATLAB. It begins with data pre-processing step. The companies that do not have enough amounts of closing prices are deleted from the company list.

[Table 3.1] Table of Number of Companies contained in the experiment for specific market index. 'Total' means the total number of components (companies) in the index. 'Used' means the number of companies left after the data pre-processing step. 'Deleted' means the number of companies deleted in the data pre-processing step. 'Winner'/'Loser' means the number of companies are labelled with 'Winner'/'Loser'.

| Index Name | Total | Used | Deleted | Winner | Loser |
|---|---|---|---|---|---|
| FTSE100 | 101 | 98 | 3 | 32 | 32 |
| DAX | 30 | 30 | 0 | 10 | 10 |
| HANGSENG | 50 | 49 | 1 | 16 | 16 |
| NASDAQ | 100 | 100 | 0 | 33 | 33 |

From [Table 3.1], the number of deleted companies is relatively small compared to the total number of components. The next step is to apply the '1/3 Average Price' method for the remaining companies, labeled them with 'winner' if the corresponding average prices are the largest 1/3 of the sequence of average prices, and 'loser' if the average prices are the last 1/3 of the sorted sequence. The result data is generated by joining the 'winner' and 'loser' companies in matrix form vertically. In this matrix each column represents each date and each row represents each company. The daily log-return matrix is computed from the joint matrix. This log-return matrix is then used for Leave-One-Out Cross-Validation of 1-NN algorithm.

*3.1. Analysis of total error and separate error*

The error of Leave-One-Out Cross-Validation for 1NN is performed for different time periods. (i.e. from 3 months to 18 months) The total error is referred to the number of misclassified points for both 'winner' and 'loser' companies.[Figure 3.2] shows the results of total error analysis for different indices. The errors generated using different functions of measurement are almost identical. The proximity can generate a bit smaller error than the distance function for several months. The total error for time period of 3 months is the minimum for three indices. For indices FTSE, DAX and NASDAQ, the errors are mostly around 50%.This means the prices are random and there is no sign of possibility of prediction. For HANGSENG index the error percentage are less than 50%.This shows that this market is not completely random. To analyze the characteristics for 'winner' and 'loser' companies, the separate error analysis is applied. The sum of 'winner' error number and 'loser' error number should be the same as the total error number.

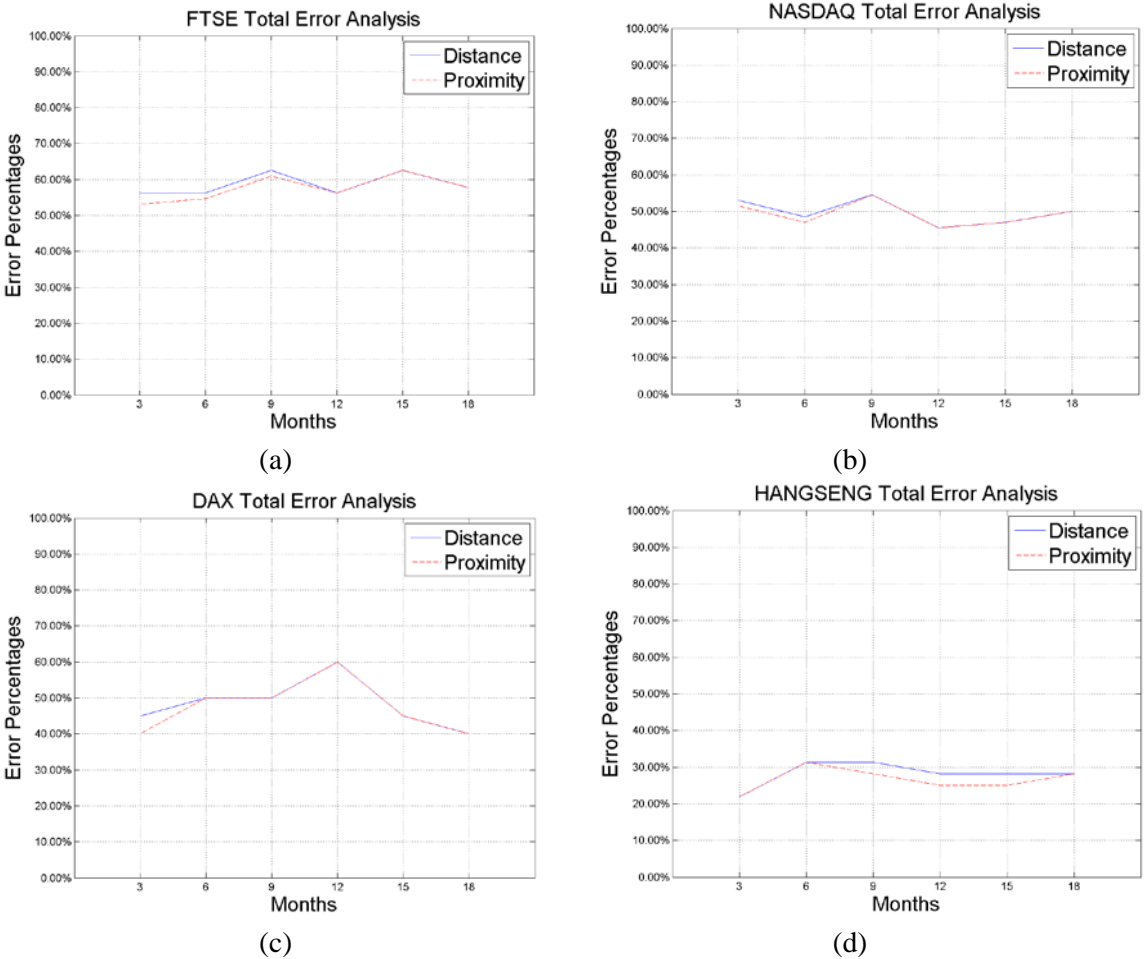

(a)     (b)
(c)     (d)

[Figure 3.2] Figures of total error analysis of Leave-One-Out Cross-Validation for 1-NN results from 3 months initial time period to 18 months initial time period for different markets (a) FTSE index (b) DAX index (c) HANGSENG index (d) NASDAQ index

The result of this analysis is show in [Figure 3.3]. For indices FTSE and NASDAQ, the error percentages are approximately 50% for both 'winner' and 'loser' companies. Hence the separate error analysis shows that there is no sign of possibility of prediction. For DAX index, the errors are ranged from 30% to 60%, it can be considered as a border case and the 'loser' companies in general have slightly lower errors than 'winner' companies. For HANGSENG index, both errors are below 50% and the errors of 'winner' companies are much smaller than the errors of 'loser' companies. Hence this means for HANGSENG index, there are some conclusions about predictability for the 'winner' companies.

*3.2. Visualization using histograms of correlation distance matrix*

The results from [Figure 3.2] and [Figure 3.3] shows the minimum error rate occurs when the time interval is within 3 months for FTSE, DAX and HANGSENG index. The sign of possibility of prediction in this time period is most dominant among all. The histograms of correlation distances can be used to study the distribution of correlation distance. For each index, a set of four histogram plots are generated. They represent the distributions of in-class (winner/winner or loser/loser) correlation distances and cross-class (winner/loser) correlation distances. For [Figure 3.4], [Figure 3.5], they are histograms for FTSE and NASDAQ indices. They have no signs of possibility of prediction.

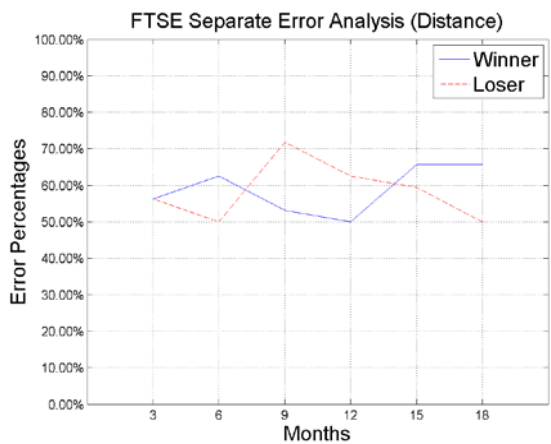

(a)

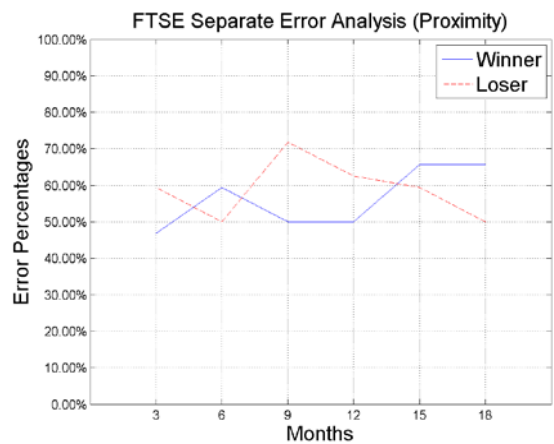

(b)

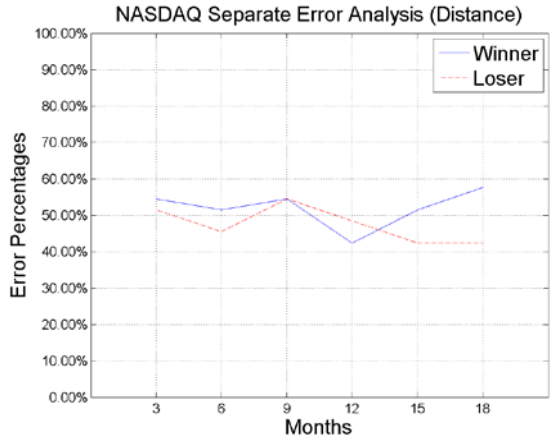

(c)

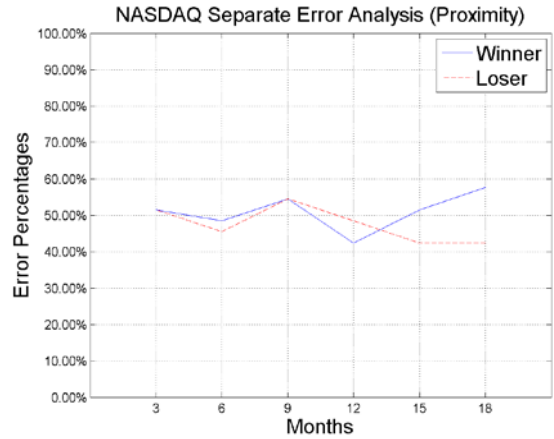

(d)

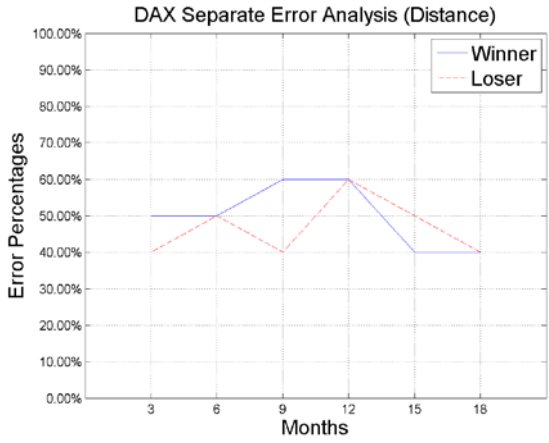
(e)
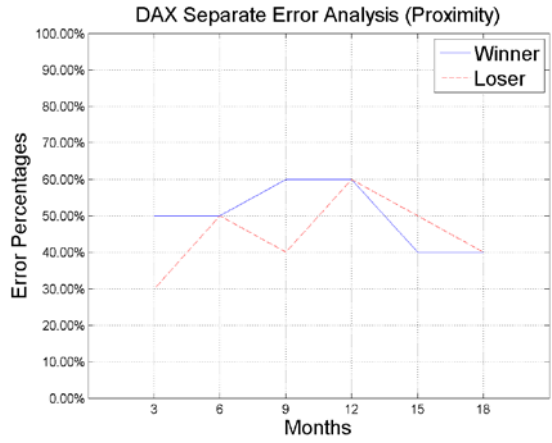
(f)
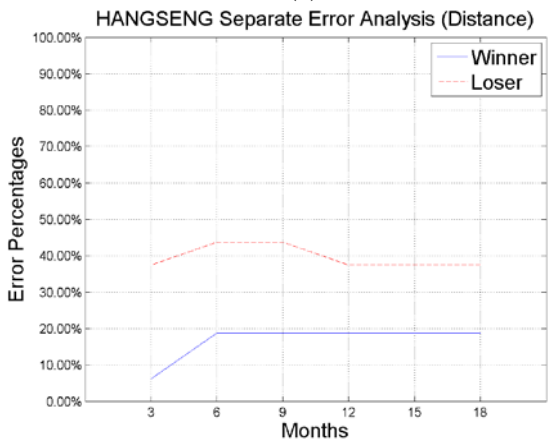
(g)
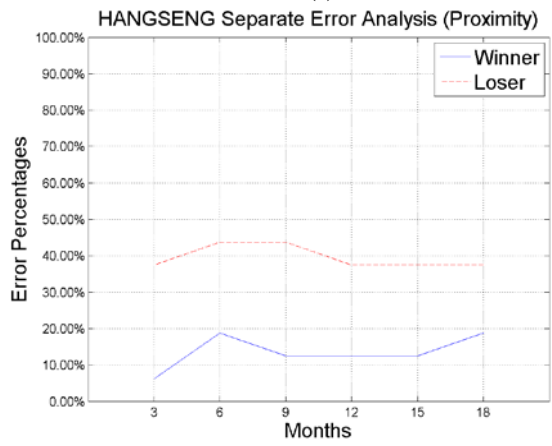
(h)

[Figure 3.3] Figures of separate error analysis of Leave-One-Out Cross-Validation for 1-NN results from 3 months initial time period to 18 months initial time period for different markets

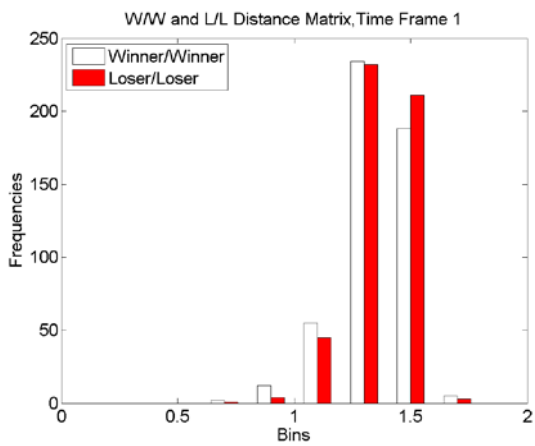
(a)
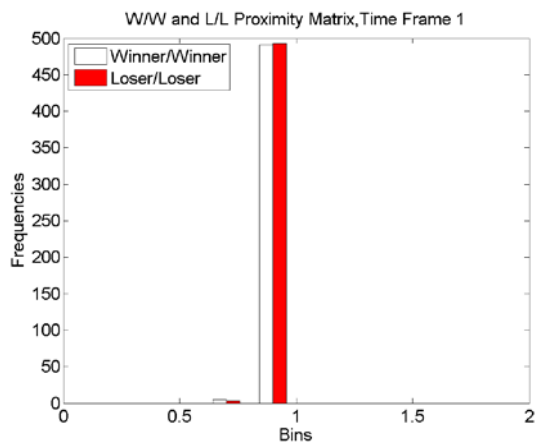
(b)

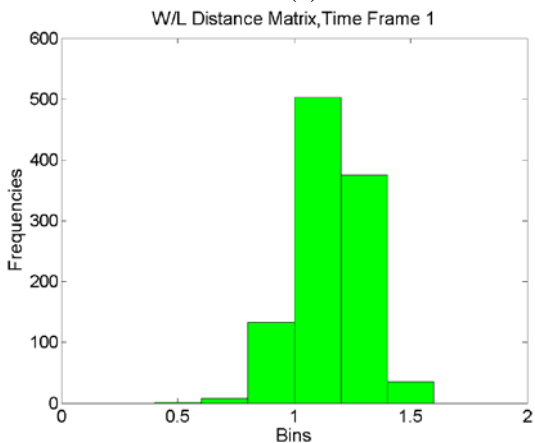
(c)
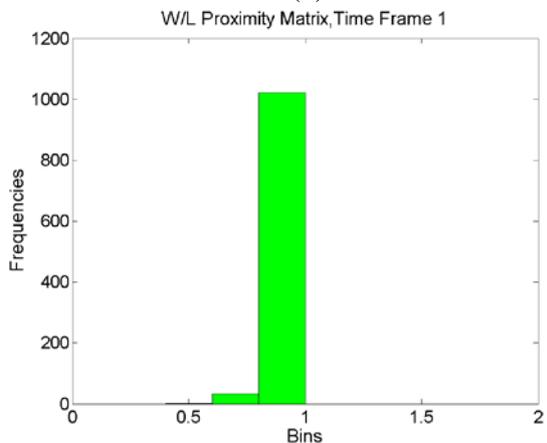
(d)

[Figure 3.4] Histograms of different expressions of correlation distances for FTSE index when using first 3 months closing prices for experiment. (a) Using Distance between winner/winner and loser/loser companies (b)Using Proximity between winner/winner and loser/loser companies (c) Using Distance between winner and loser companies (d) Using Proximity between winner and loser companies

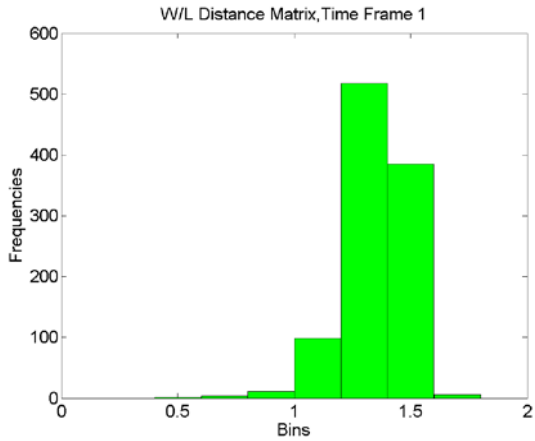
(a)
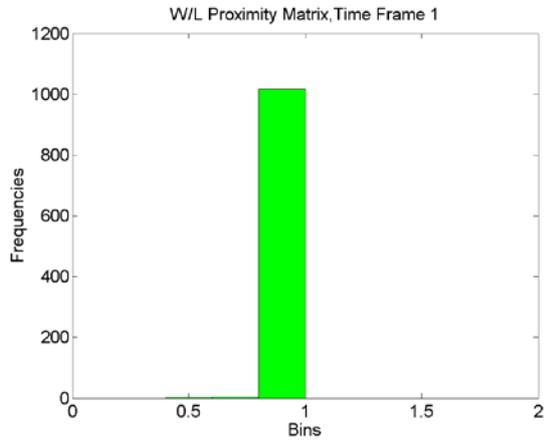
(b)
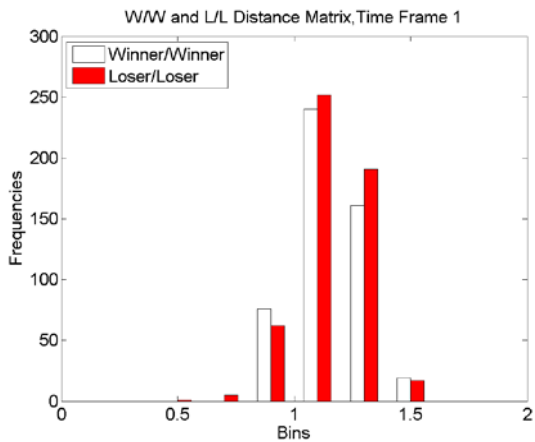
(c)
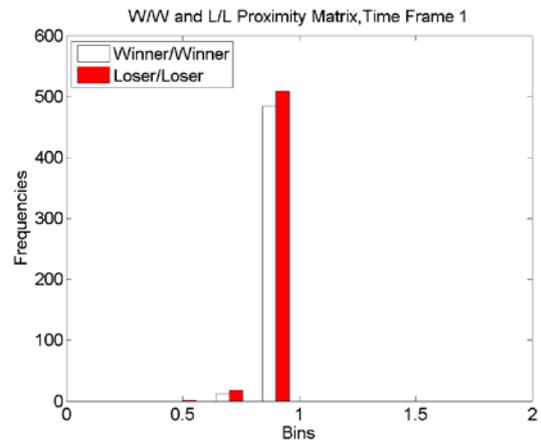
(d)

[Figure 3.5] Histograms of different expressions of correlation distances for NASDAQ index when using first 3 months closing prices for experiment. (a) Using Distance between winner/winner and loser/loser companies (b)Using Proximity between winner/winner and loser/loser companies (c) Using Distance between winner and loser companies (d) Using Proximity between winner and loser companies

The figures are the results of LOOCV error analysis for 'winner' and 'loser' companies separately. For the figures on the same row, these represent the result of experiment using companies from the same market but different expressions of distance (use Distance or Proximity). (a)FTSE index use Distance expression (b)FTSE index with Proximity expression (c)DAX index use Distance expression (d)DAX index with Proximity expression (e)HANGSENG index use Distance expression (f) HANGSENG index with Proximity expression (g)NASDAQ index use Distance expression (h) NASDAQ index with Proximity expression

The in-class histograms show that the distribution of distances between 'winner' themselves and 'loser' themselves are almost identical. The distributions of in-class distances and cross-class distances are quite similar as well. This makes 'winner' and 'loser' companies are not easily separated. Since k-NN is structure sensitive algorithm, if the points are mixed together, it generates errors. Hence if 'winner' and 'loser' companies are mixed together, this market has no sign of possibility of prediction. [Figure 3.6] is the histogram of correlation distances for DAX index.

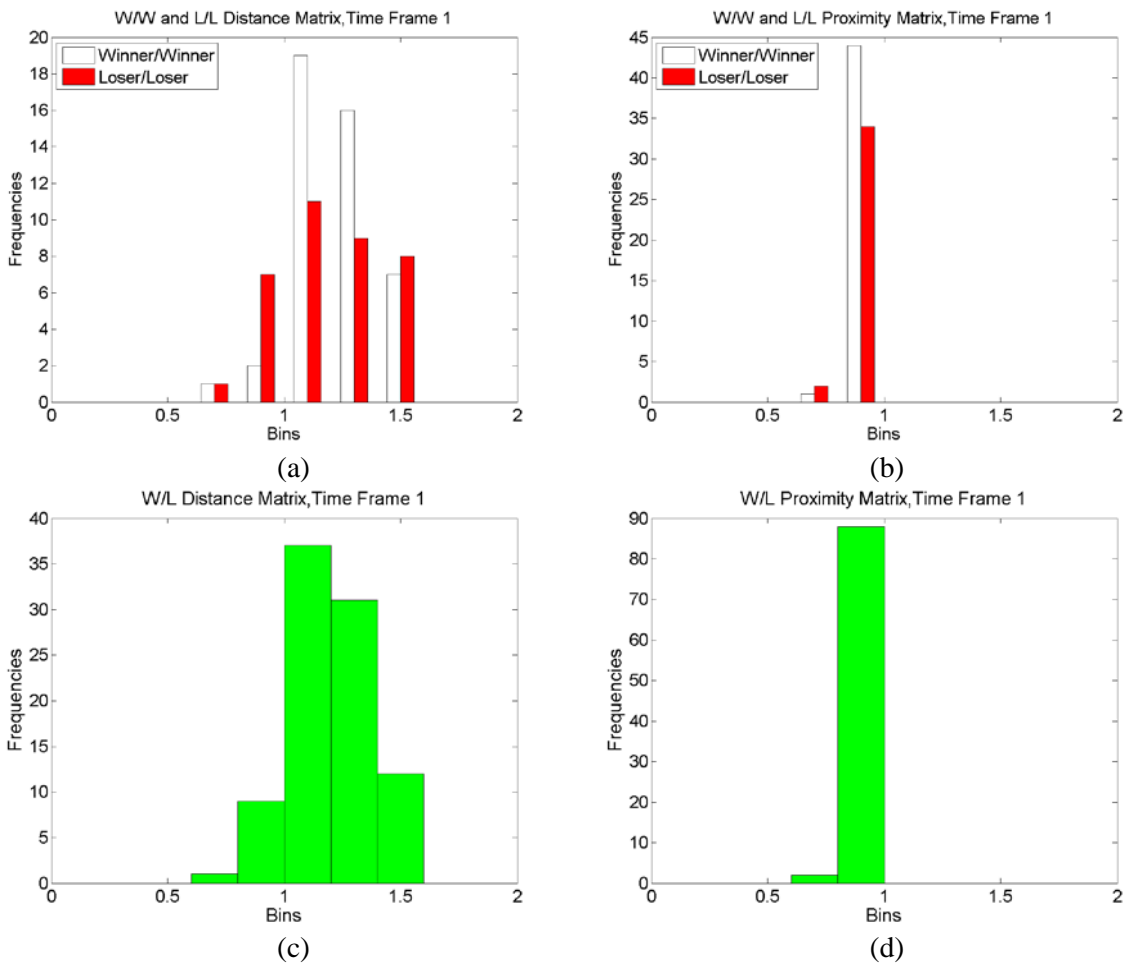

(a)  (b)
(c)  (d)

[Figure 3.6] Histograms of different expressions of correlation distances for DAX index when using first 3 months closing prices for experiment. (a) Using Distance between winner/winner and loser/loser companies (b)Using Proximity between winner/winner and loser/loser companies (c) Using Distance between winner and loser companies (d) Using Proximity between winner and loser companies

The DAX index has some possibility of prediction but this sign is not very clear. From the in-class distributions, the 'winner' and 'loser' seems to be separated, but this separation is not clear enough since the distribution peak of winner/winner distances is slightly shifted to the left hand side of the distribution peak of loser/loser distances. The histograms [Figure 3.7] show the distributions of correlation distances for HANGSENG index. For this index, it has a clear sign of possibility of prediction. The distribution of winner/winner distances is on the left hand side of the distribution of loser/loser distances. Hence the 'winner' companies are more compactthan 'loser' companies and there is a good separation between 'winner' and 'loser' companies.

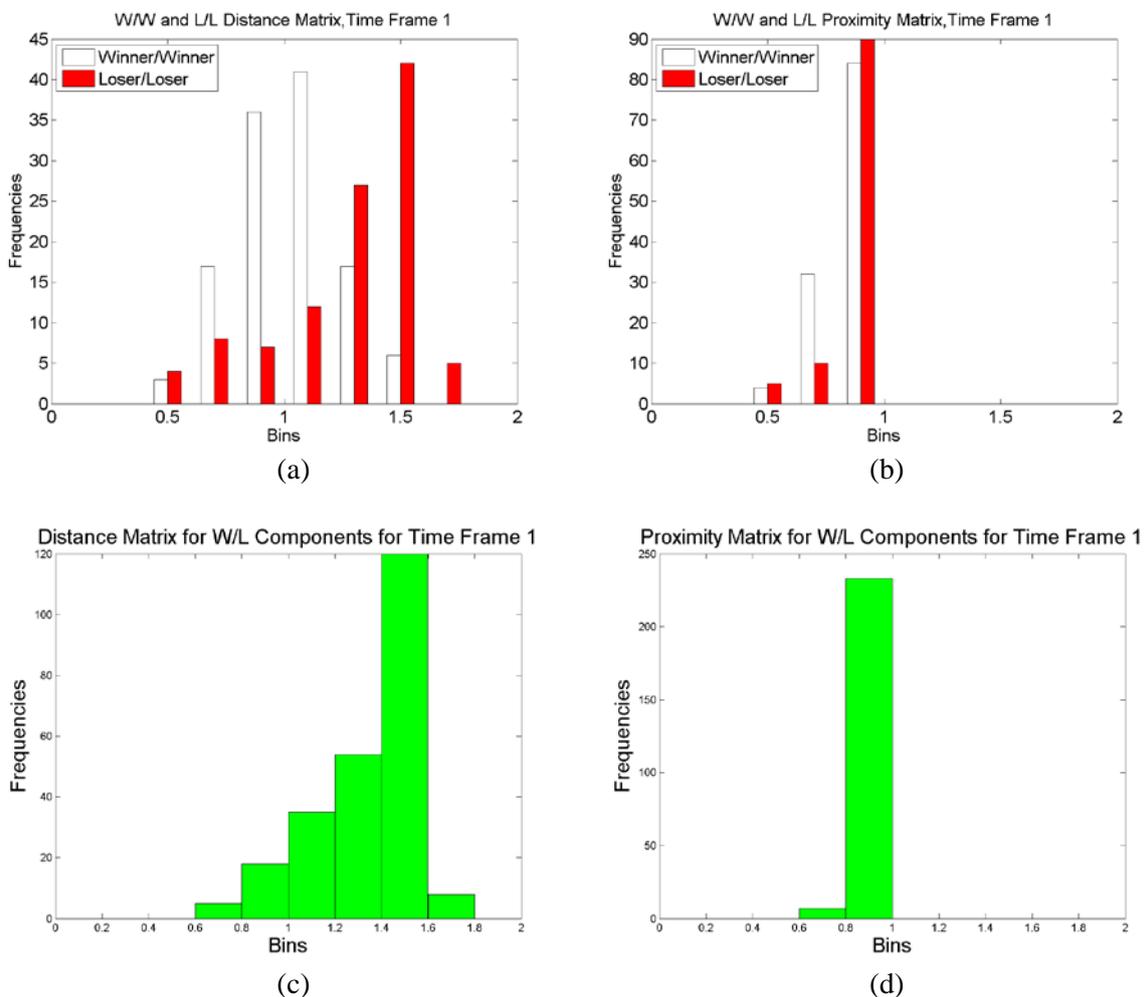

(a) (b)

(c) (d)

[Figure 3.7] Histograms of different expressions of correlation distances for HANGSENG index when using first 3 months closing prices for experiment. (a) Using Distance between winner/winner and loser/loser companies (b)Using Proximity between winner/winner and loser/loser companies (c) Using Distance between winner and loser companies (d) Using Proximity between winner and loser companies

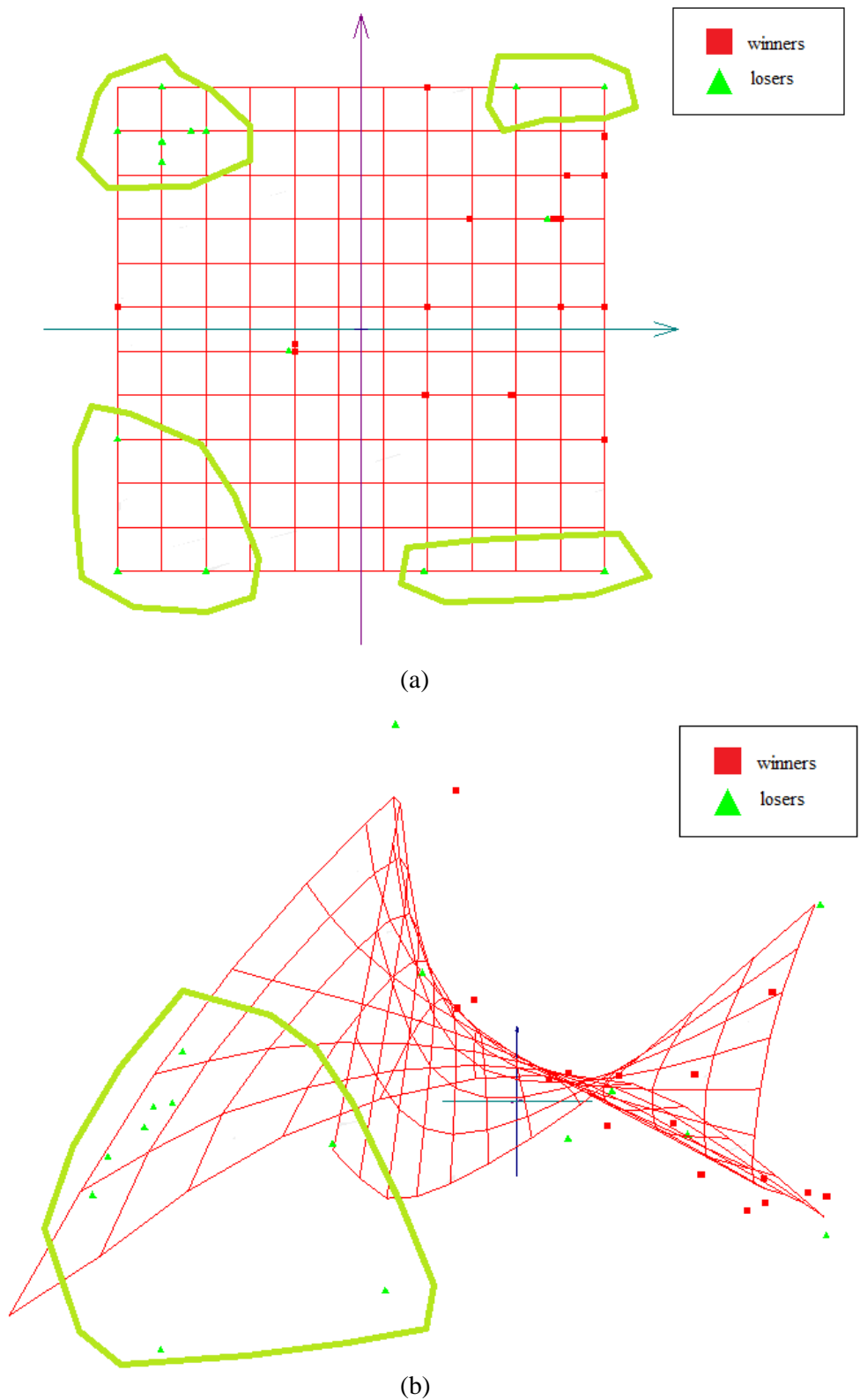

[Figure3.8] Visualization of components (companies) of HANGSENG index (log-returns) using elastic maps: (a) 2D - Elastic Map (b) 3D - Principal Manifold Graph The Hand-made green lines show the "clusters" of loser companies

*3.3. Visualization of using 2-D/3-D elastic maps*

The figures of companies are always good idea to visualize this characteristic. The most difficult part of plotting is the dimension of the dataset. For this data, the dimension is relatively high. (i.e. each date is a dimension) The principal graphs and manifolds can be used to produce the plots with lower dimension. In previous studies, the metaphor of elastic membrane and plate is used to construct one-, two- or three dimensional principal manifold approximations of various topologies. The mean squared distance approximation error and the elastic energy of the membrane together formed a functional to be optimized [15]. This idea of using elastic graphs is demonstrated on several practical examples: from comparative political science, data analysis in molecular biology and analysis of dynamical systems for biochemical modeling [16]. The software 'ViDaExpert' [17] developed by Dr. Andrei Zinovyev uses this idea of elastic energy to compute elastic map and net using the principal manifolds. This software enables users to visualize multidimensional data with the idea of using principal object to reduce the dimension of this data.

In our experiment, we use this software to visualize the log-return prices for all companies within a time period of the initial first 3 months. There are two parameters of this elastic map algorithm to generate the graph, the coefficients of 'stretching elasticity' $\lambda$ and the 'bending elasticity' $\mu$. To have the good performance approximation for the principal manifold, I fixed $\lambda \approx 0$ and $\mu \approx 8.1$. For each index, two types of graphs are generated using ViDaExpert. The first one is the "2D -Elastic Map". It displays the estimation of 2D density of companies in the internal manifold coordinates. The second graph is the "3D - Principal Manifold Graph", it displays the companies in the first three principal component coordinates.

For HANGSENG index with first 3-month time frame, the elastic maps [Figure 3.8] are generated. From (a), it shows that the winners and losers are almost separated nicely. The losers generate 4 "clusters" within the internal coordinates. These "clusters" are located on each corner of the coordinate plane. From (b), it shows the winners are closer to each other than the loser companies as the red points are more compact to each other. However, the map boundary is not linear since some loser companies that are very close to the winner companies and there exists around six points that is seems to be very close to the red points. These green points can be considered as outliers. As the error of k-NN is dominated by the structure of data, this visualization supports the LOOCV error analysis.

For DAX index, the elastic maps [Figure 3.9] have a bit worse results than the one for HANGSENG index. From (a), the losers generate 2 main "clusters" one near the top left corner and another one on the right hand side. It is still possible to separate the 'winner' and 'loser' companies but it is not as clear as the result for HANGSENG index. From (b), it seems that the 'winner' company are close to each other in the middle of the map while the losers formed two "clusters" on the bottom. Both 'winner' and 'loser' companies have similar closeness between their own points. Hence from the figure, it is different to the k-NN analysis that there is still a small amount of possibility of predicting the future "success".

For the elastic maps of companies of FTSE index [Figure 3.10] and NASDAQ index [Figure 3.11], there is no sign of separation. There is a small cluster of 'loser' companies for FTSE index but there are no clusters for 'winner' companies. For NASDAQ, there is no clear sign of clusters in the internal coordinate maps. The elastic maps give negative results for the companies of these two indices. The k-NN classifier is structure sensitive the elastic maps support the LOOCV error analysis. Therefore the experiments have negative results and there is no evidence showing the possibility of predicting future success.

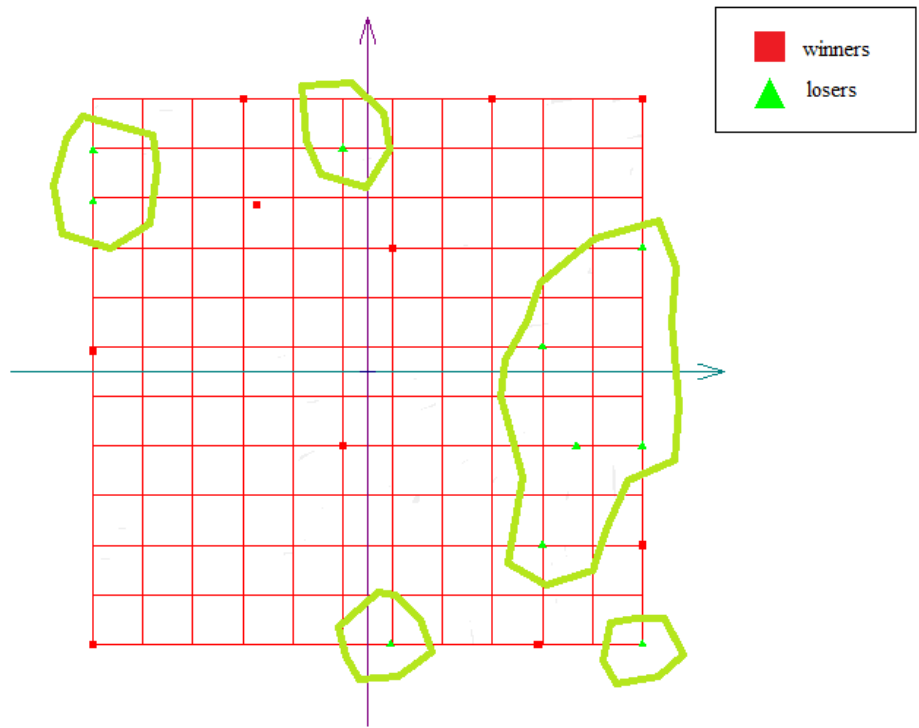

(a)

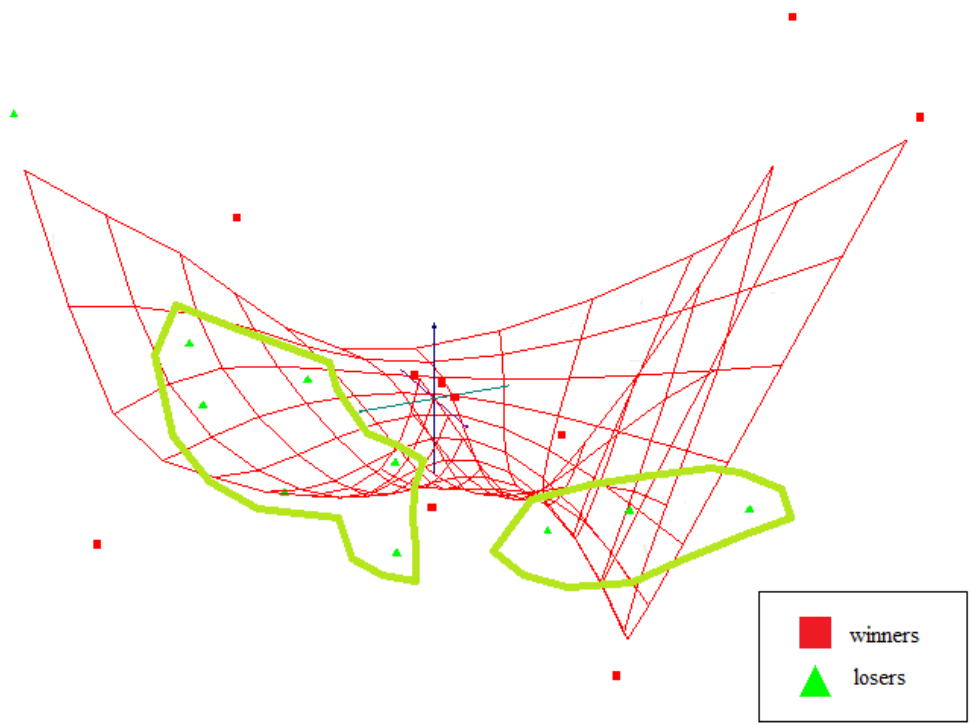

(b)

[Figure3.9] Visualization of components (companies) of DAX index (log-returns) using elastic maps. (a) 2D - Elastic Map (b) 3D - Principal Manifold Graph The Hand-made green lines show the "clusters" of loser companies

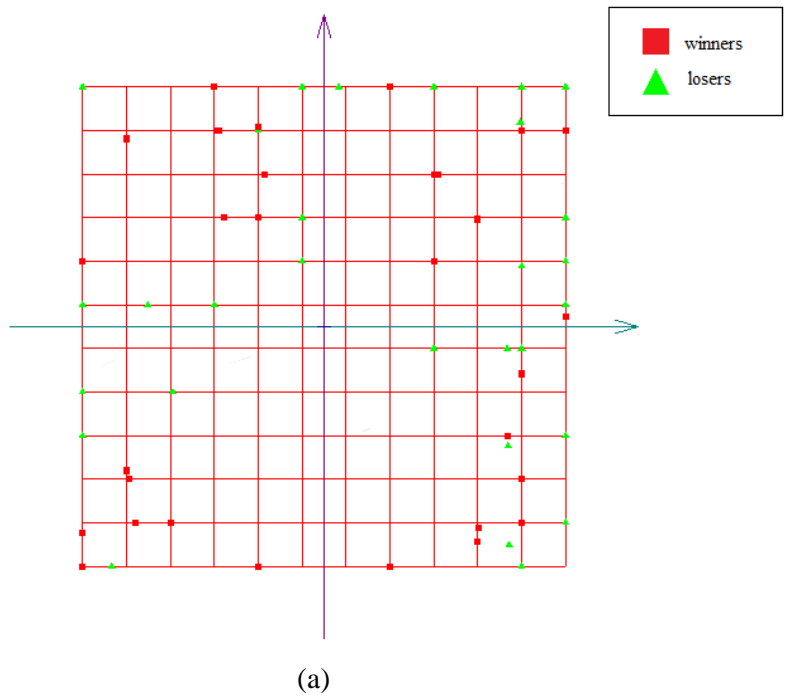

(a)

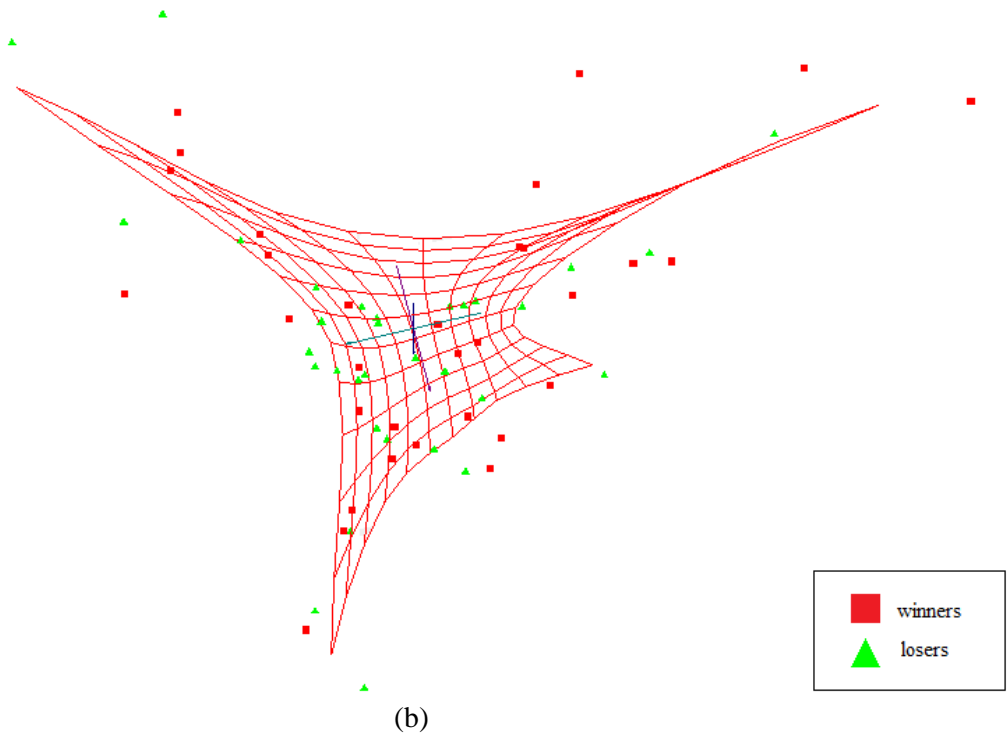

(b)

[Figure 3.10] Visualization of components (companies) of FTSE index (log-returns) using elastic maps. (a) 2D - Elastic Map (b) 3D - Principal Manifold Graph

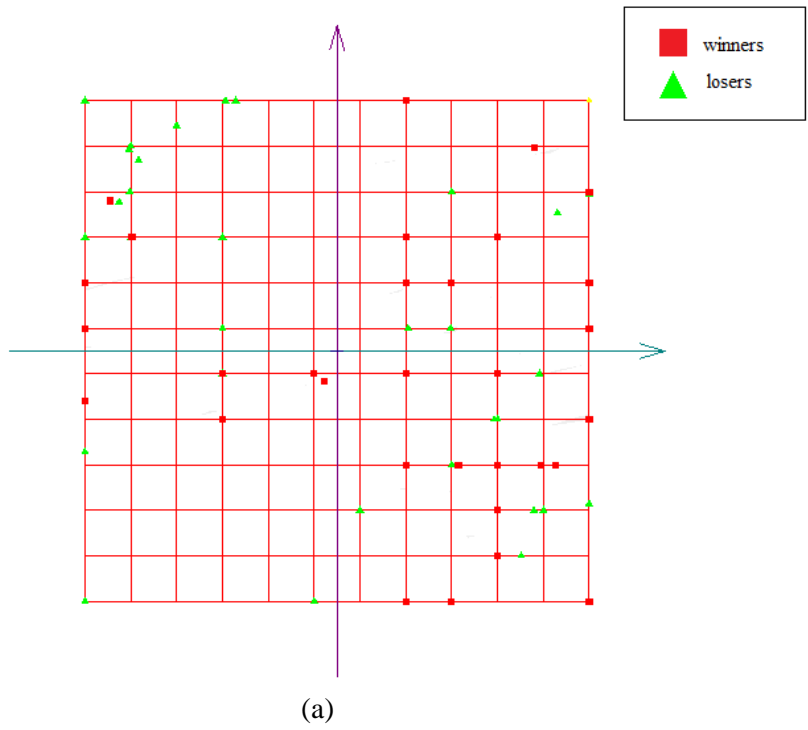

(a)

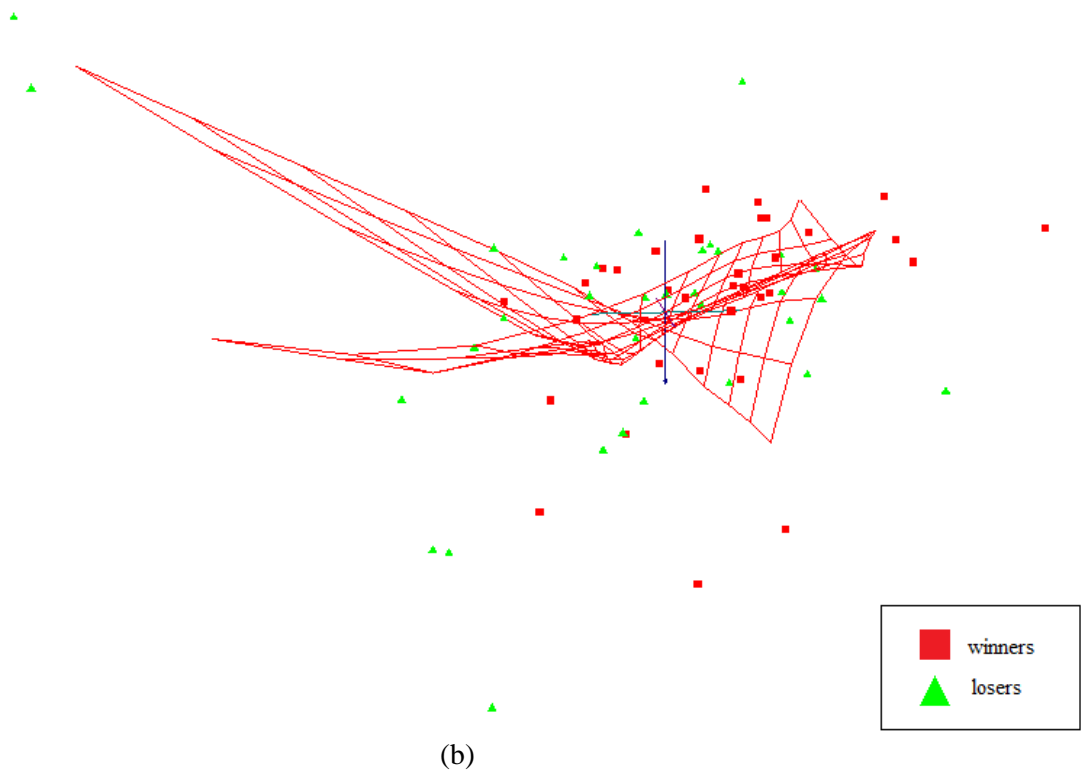

(b)

[Figure 3.11] Visualization of components (companies) of DAX index (log-returns) using elastic maps. (a) 2D - Elastic Map (b) 3D - Principal Manifold Graph

*3.4. Analysis using proportion estimate*

Consider the error to be the probability of occurrence event, the sample distribution of error is computed by applying proportion estimate.

[Table 3.12] Table of result using proportion estimate analysis for 'winner' and 'loser' companies using Distance/ Proximity. μ is the sample distribution mean, and σ is the sample distribution standard deviation.

Distance

| Winner | M | σ | Loser | μ | σ |
|---|---|---|---|---|---|
| 3 months | 0.0625 | 0.0605 | 3 months | 0.3750 | 0.1210 |
| 6 months | 0.1875 | 0.0976 | 6 months | 0.4375 | 0.1240 |
| 9 months | 0.1875 | 0.0976 | 9 months | 0.4375 | 0.1240 |
| 12 months | 0.1875 | 0.0976 | 12 months | 0.3750 | 0.1210 |
| 15 months | 0.1875 | 0.0976 | 15 months | 0.3750 | 0.1210 |
| 18 months | 0.1875 | 0.0976 | 18 months | 0.3750 | 0.1210 |

Proximity

| Winner | μ | σ | Loser | μ | σ |
|---|---|---|---|---|---|
| 3 months | 0.0625 | 0.0605 | 3 months | 0.3750 | 0.1210 |
| 6 months | 0.1875 | 0.0976 | 6 months | 0.4375 | 0.1240 |
| 9 months | 0.1250 | 0.0827 | 9 months | 0.4375 | 0.1240 |
| 12 months | 0.1250 | 0.0827 | 12 months | 0.3750 | 0.1210 |
| 15 months | 0.1250 | 0.0827 | 15 months | 0.3750 | 0.1210 |
| 18 months | 0.1875 | 0.0976 | 18 months | 0.3750 | 0.1210 |

[Table 3.12] shows the results of proportion estimate analysis. In general, the 'loser' companies have sampling standard deviation of 0.1210 >0.1 where the sampling standard deviation for 'winner' companies are approximately 0.0900 <0.1. The results using different correlation distances do not make much difference. There is no overlap in the interval between 'winner' and 'loser' companies.

**4. Conclusions**
In this case study, a backward analysis is used as the principal idea of experiments. The past and future data are used for labeling but only the past data is used in k-NN classification. From the experiments, the results show that in this specific period of time, there is a phenomenon (i.e. the winner companies are closer to each other) for the HANGSENG index which means that there could be some conclusion about predictability of long term success of companies in the HANGSENG index. Using different expressions of distances do not make much difference in the k-NN LOOCV error analysis. For the only index with positive result, the more closing prices given (i.e. the longer initial time frame for initial information) does not improve the predictability for the k-NN predictor since the LOOCV error analysis does not decrease when time period is larger.

For experiment results of DAX index, the error analysis rejects the possibility of future prediction. However, the for the first 3-month experiment, the LOOCV error is just a bit below than 50%. Therefore this is the boundary case for our experiment and it is the reason why it gives different results in the visualization result of elastic map. In this case, it can be concluded that there is still a small amount of possibility of future success prediction. For FTSE and NASDAQ indices, it can be concluded that the long term success is not predictable on the base of daily closing prices in the initial time frame as the LOOCV error analysis is higher than 50%.

The positive result can also indicate for HANGSENG index, the past prices have information about future prices. Hence there is a high possibility that technical analysis is profitable for this case. Since HANGSENG index is the index of developing market, this success in predictability can show that for index of developing markets, the long term success is more predictable. Technical analysis may be applied these young markets since for HANGSENG and DAX index, the historical prices may have some information. For FTSE and NASDAQ index, the prices have less information about the long term success.